\documentclass[10pt]{article}
\usepackage{spconf,amsmath,graphicx}
\usepackage{amssymb,amsfonts}
\usepackage{bm}
\usepackage{url}
\usepackage{xcolor}
\usepackage{hyperref}
\hypersetup{hidelinks,
	colorlinks=true,
	allcolors=black,
	pdfstartview=Fit,
	breaklinks=true}
\usepackage[font=small,labelsep=period]{caption}
\usepackage{bm}
\usepackage{multirow}
\usepackage{booktabs}
\usepackage{makecell}
\usepackage{bbding}
\usepackage{balance}
\usepackage{float}
\usepackage{setspace}

\title{Sub-band and full-band Interactive U-Net with DPRNN \\
for Demixing Cross-talk Stereo Music}
\name{
Han Yin{\rm\textsuperscript{1}}, Mou Wang{\rm\textsuperscript{2}}, Jisheng Bai{\rm\textsuperscript{1,3}}, 
Dongyuan Shi{\rm\textsuperscript{3}}, Woon-Seng Gan{\rm\textsuperscript{3}}, Jianfeng Chen{\rm\textsuperscript{1}}  
}

\address{
        \textsuperscript{1} School of Marine Science and Technology, Northwestern Polytechnical University, Xi'an, China\\
        \textsuperscript{2} Institute of Acoustics, Chinese Academy of Sciences, Beijing, China\\
        \textsuperscript{3} School of Electrical \& Electronic Engineering, Nanyang Technological University, Singapore
}

\begin{document}
%
\maketitle
\begin{abstract}
This paper presents a detailed description of our proposed methods for the ICASSP 2024 Cadenza Challenge.
Experimental results show that the proposed system can achieve better performance than official baselines.
\end{abstract}
\section{Introduction}
\label{sec:intro} 
How to provide hearing-impaired people with a better listening experience is a challenging problem.
One approach is to develop a signal processing system that allows a personalized rebalancing of the music.
The ICASSP 2024 Cadenza Challenge \cite{ICASSP2024-Cadenza} encourages participants to separate music into different tracks (vocals, bass, drums and other), 
and then intelligently remix the tracks in a personalized manner to enhance the listening experience for people with hearing loss.


In this paper, we propose a sub-band and full-band interactive U-Net model to demix the cross-talk stereo music.
The model extracts dynamic audio information from the sub-bands and the full-band respectively, and then uses convolutional layers to adaptively fuse the embeddings from different branches. 
Furthermore, a neural beamformer composed of multilayer perceptrons (MLPs) is applied to attenuate the effect of cross-talk. 

\section{Methods}
\subsection{Proposed Demixing Model}

Fig. \ref{fig:model1} shows the overall pipeline of the proposed demixing model.
Short-time Fourier transform (STFT) is applied on the original music to get the full-band spectrogram, denoted as $\bm{X} \in \mathbb{C}^{C \times T \times F }$, where $C$, $T$ and $F$ are the number of channels, frames and frequency points respectively.
The full band is divided into two predefined sub-bands, denoted as $\bm{X}_1 \in \mathbb{C}^{C \times T \times F_1 }$ and $\bm{X}_2 \in \mathbb{C}^{C \times T \times F_2 }$,
where $F_1$ and $F_2$ are predefined sub-band ranges.

$\bm{X}$, $\bm{X}_1$ and $\bm{X}_2$ are fed into three U-Nets with dual-path recurrent neural network (DPRNN) \cite{luo2020dual} to extract dynamic audio information, an interactive block is then applied to fuse the masks generated by different branches.
Furthermore, MLPs in the neural beamformer are used to adaptively adjust the phase of the estimated complex-valued mask to mitigate the effect of cross-talk on the listening experience.

\subsubsection{U-Net with DPRNN}
\begin{figure}
\centering
\centerline{\includegraphics[width=0.45\textwidth]{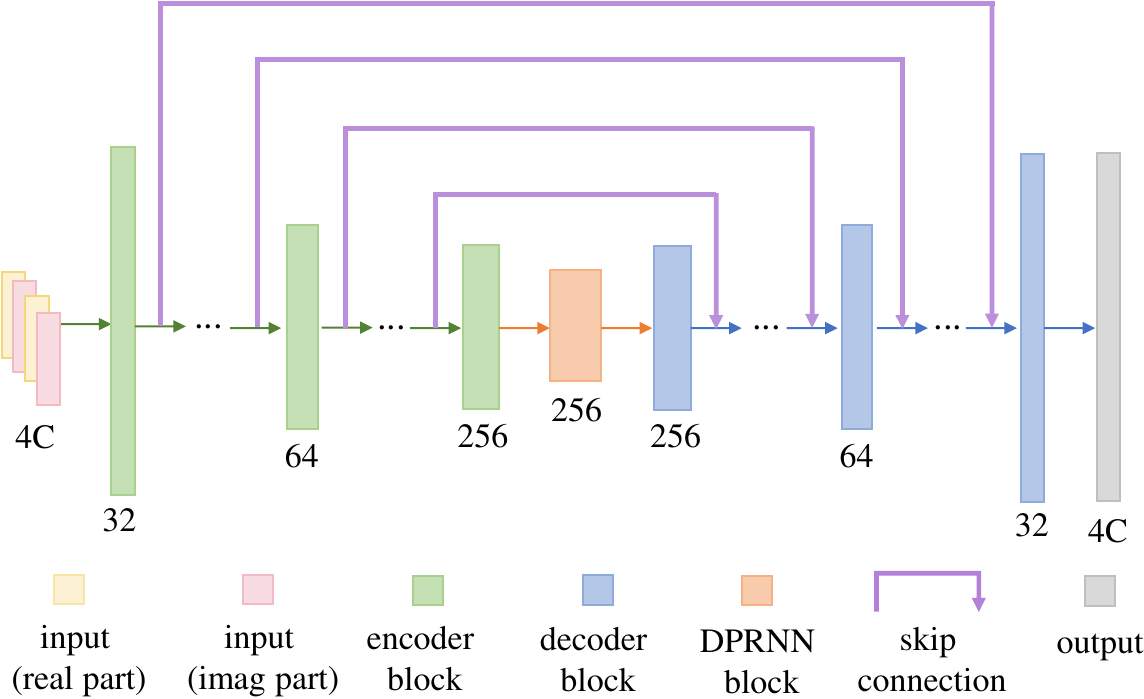}}
\caption{The architecture of the U-Net with DPRNN}
\label{fig:unet}
\end{figure}

As shown in Fig. \ref{fig:unet}, this module consists of encoder blocks, a DPRNN block and decoder blocks.
We use two-dimensional convolutional layers in the encoder block to extract features from the original spectrogram.
The DPRNN block, which consists of two bi-directional LSTMs with a residual connection, is used to alternately model the time and frequency information.
In decoder blocks, we use two-dimensional deconvolution layers to restore the time-frequency domain mask.

\subsubsection{Interactive Block}
\begin{figure*}
\centering
\centerline{\includegraphics[width=0.9\textwidth]{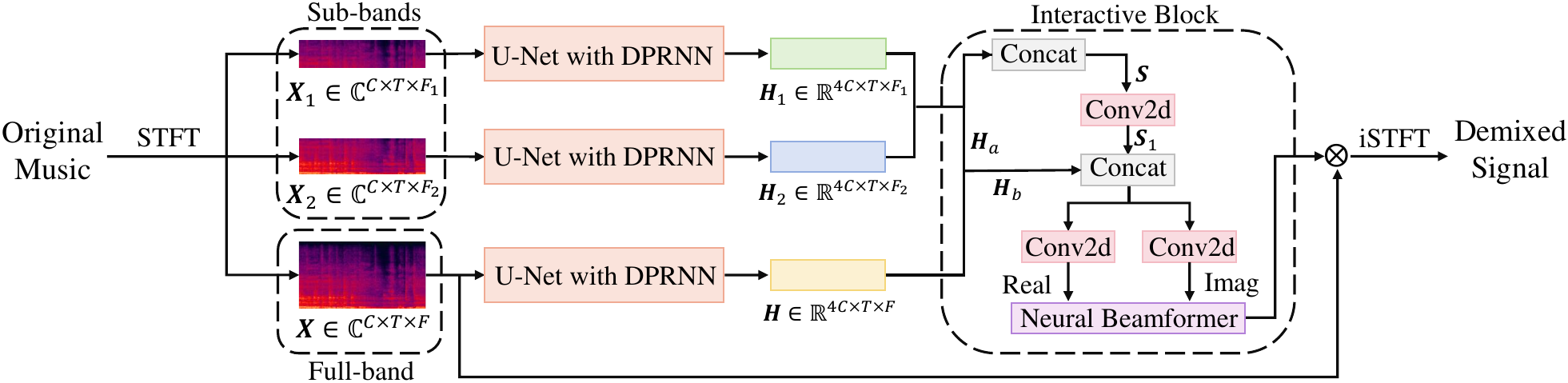}}
\caption{The architecture of the proposed sub-band and full-band interactive U-Net with DPRNN for demixing cross-talk stereo music}
\label{fig:model1}
\end{figure*}

As shown in Fig. \ref{fig:model1}, the sub-bands and the full band are separately passed through a UNet with DPRNN to generate the corresponding masks, denoted as $\bm{H}_1\in\mathbb{R}^{4C \times T \times F}$, $\bm{H}_2\in\mathbb{R}^{4C \times T \times F_1}$ and $\bm{H}\in\mathbb{R}^{4C \times T \times F_2}$.

In the proposed interactive block, the full band mask is firstly divided into two parts, as formulated in:
\begin{equation}
\begin{aligned}
    \bm{H}^a &= \bm{H}[:,:,:F_1+F_2]\\
    \bm{H}^b &= \bm{H}[:,:,F_1+F_2:]
\end{aligned}
\end{equation}
Then, $\bm{H}_1$, $\bm{H}_2$ and $\bm{H}^a$ are concatenated to produce a new sub-band mask $\bm{S} \in \mathbb{R}^{8C \times T \times (F_1+F_2)}$, which is fed into a convolutional layer, resulting in $\bm{S}_1 \in \mathbb{R}^{4C \times T \times (F_1+F_2)}$. 
$\bm{H}^b$ and $\bm{S}_1$ are concatenated to generate a new full-band mask, which is passed through two convolutional layers to estimate 
the real and imaginary parts of the complex-valued mask.
Finally, we pass the real part and the imaginary part of the estimated mask through an MLP in the frequency domain respectively for adaptive adjustment in the neural beamformer.
The kernel size in all convolutional layers is set to 1.

\label{sec:methods}

\subsection{Loss Function}
In the proposed system, the loss function is formulated as:
\begin{equation}
\begin{aligned}
    Loss &= |\bm{y},\bm{\hat{y}}|+|\bm{Y},\bm{\hat{Y}}|+|{\rm real}(\bm{Y}),{\rm real}(\bm{\hat{Y}})|\\
    &+|{\rm imag}(\bm{Y}),{\rm imag}(\bm{\hat{Y}})|
\end{aligned}
\end{equation}
where $\bm{y}$ and $\bm{\hat{y}}$ are the estimated signal and reference signal respectively, $\bm{Y}$ and $\bm{\hat{Y}}$ are corresponding spectrograms.

\section{Experiments and Discussion}
Adam optimizer is used and the initial learning rate is set to 0.001.
Batch size is set to 2 and the number of gradient accumulations is 6.
The parameter size of the proposed model is 152.4 MB. 
Computational resources and configurations of the sub-bands are presented in Table \ref{tab:freq}.

Table \ref{tab:eval} shows the HAAQI results of different systems on the evaluation set and the validation set respectively.
For each track, we train the proposed model multiple times and obtain the top 3 models. 
``Ensemble" is the weighted average result of the three models, while no ensemble represents the result generated by the best model.
Results show that the proposed model can achieve HAAQI scores comparable to HDmucs \cite{defossez2021hybrid}, and the ensemble can further improve the performance.

\begin{table}[htbp]\small
\centering
\setlength{\abovecaptionskip}{0cm} 
\caption{Configuration of sub-bands and corresponding GPU memories for training (Input audio: 4s, FFT size: 2048, Hop size: 600)}
\renewcommand\arraystretch{1}{
\setlength{\tabcolsep}{2.3mm}{
\begin{tabular}{c|cccc|c}
\hline
 \multirow{2}{*}{Track}                         & \multicolumn{2}{c}{Sub-band 1} & \multicolumn{2}{c|}{Sub-band 2} & Training\\
                      & Start           & End       & Start   & End  & GPU memory\\
\hline
Vocals             &   0 kHZ         & 4 kHz & 4 kHz & 10 kHz & 9.52 GB\\
Drums        &   0 kHZ         & 6 kHz & 6 kHz & 10 kHz & 9.49 GB\\
Bass              &   0 kHZ         & 1 kHz & 1 kHz & 6 kHz & 9.23 GB\\
Other             &   0 kHZ         & 7 kHz & 7 kHz & 11 kHz & 9.41 GB\\
\hline
\end{tabular}
}
}
\label{tab:freq}
\end{table}

\begin{table}[htbp]\small
\centering
\setlength{\abovecaptionskip}{0cm} 
\caption{HAAQI Results of different models on the evaluation set and validation set}
\renewcommand\arraystretch{1}{
\setlength{\tabcolsep}{1mm}{
\begin{tabular}{c|ccc|ccc}
\hline
\multirow{2}{*}{System}                    & \multicolumn{3}{c|}{Evaluation Set} & \multicolumn{3}{c}{Validation Set}\\
                                          & Left       & Right  & Overall  & Left       & Right  & Overall\\
\hline
OpenUnmix \cite{stoter2019open}                & 0.507 & 0.515 & 0.511  & 0.599 & 0.594 & 0.596\\
HDemucs \cite{defossez2021hybrid}              & 0.566 & 0.574 & 0.570  & 0.669 & 0.667 & 0.668\\
Proposed                                  & 0.564 & 0.572 &  0.568  & 0.667 & 0.664 & 0.665 \\
Proposed(ensemble)                        & \textbf{0.581} & \textbf{0.589} & \textbf{0.585}   & \textbf{0.686} & \textbf{0.682} & \textbf{0.684}\\
\hline
\end{tabular}
}
}
\label{tab:eval}
\end{table}


\section{Conclusions}
This paper presents our proposed demixing system based on U-Net and DPRNN in the ICASSP 2024 Cadenza Challenge.
Experimental results show that the proposed system outperforms OpenUnmix \cite{stoter2019open} and HDemucs \cite{defossez2021hybrid}, 
achieving an overall HAAQI of 0.585 on the evaluation set.


\begin{spacing}{0.9}
\bibliographystyle{IEEEbib}
\bibliography{strings,refs}
\end{spacing}

\end{document}